\documentclass{article}

\usepackage{amsmath,amssymb,epsf,a4wide}

\newcommand{\ba}{\begin{eqnarray}}
\newcommand{\ea}{\end{eqnarray}}

\renewcommand{\kappa}{{k}}
\begin{document}
\begin{titlepage}

\begin{centering}
\begin{flushright}
hep-th/0103122
\end{flushright}

\vspace{0.1in}

{\Large {\bf A study of spacetime distortion around a scattered recoiling 
$D$-particle and possible astrophysical consequences}}

\vspace{0.4in}

{\bf Elias Gravanis and Nick E. Mavromatos } \\
\vspace{0.2in}
Department of Physics, Theoretical Physics, King's College London,\\
Strand, London WC2R 2LS, United Kingdom.

\vspace{0.4in}
 {\bf Abstract}

\end{centering}

\vspace{0.2in}

{\small We study a four-dimensional spacetime induced 
by the recoil of a D(irichlet)-particle, embeded in it, due 
to scattering by a moving string. The induced spacetime 
has curvature only up to 
a radius that depends on the energy of the incident string. 
Outside that region (`bubble') the spacetime is matched 
with the Minkowski spacetime.  
The interior of the bubble 
is consistent with the 
effective field theory obtained from strings, with non-trivial 
tachyon-like and antisymmetric tensor fields (in four dimensions 
the latter gives rise to an axion pseudoscalar field).
The tachyonic mode, however, does not represent the standard flat-spacetime 
string tachyon, but merely expresses the instability of the 
distorted spacetime. 
Due to the non-trivial matter content of the 
interior of the bubble, there is entropy production, which expresses
the fact that information is carried away by the recoil degrees of freedom.
We also demonstrate that a particle can be captured by the bubble, 
depending on the particle's impact parameter. 
This will result 
in information loss for an external asymptotic observer, 
corresponding to production of entropy 
propotional to the area of the bubble.
For the validity of our approach 
it is essential that the string length is a few orders of magnitude 
larger than the Planck length, 
which is  a typical situation encountered
in many D-brane-world models.
A very interesting feature of our model
is the emission of high-energy 
photons from the unstable bubble. 
As a result, the neighborhood of the recoiling D-particle 
defect may operate as a source of ultra-high-energy 
particles, which could reach the observation 
point if the source lies within the respective mean-free paths.
This may 
have possible non-trivial 
physical applications, e.g. in connection with 
the observed apparent ``violations'' of the GZK cutoff.}

\vspace{0.8in}
\begin{flushleft}
March 2001 
\end{flushleft}

\end{titlepage}

\section{Introduction}

It has been argued previously~\cite{recoil} that
the recoil of a D-particle embedded in a four-dimensional spacetime 
(obtained from appropriate compactification of a 
higher-dimensional stringy spacetime) results in the following 
metric:
\ba
ds^2 = \frac{{b'}^2 r^2}{t^2}dt^2 - \sum_{i=1}^{3} dx_i^2~, \qquad r^2 =
\sum_{i=1}^{3} x_i^2 
\label{metricrecoil}
\ea
The above metric is derived by demanding appropriate Liouville 
dressing 
of the corresponding non-critical string, the non-criticality 
being due to the presence of the recoil world-sheet vertex 
operators~\cite{kogan}, which are {\it relevant operators } in a world-sheet
renormalization group sense. As discussed in \cite{szabo}, the 
dimensionless 
parameter $b'$ depends  on the energy of the incident string, $E$, 
and, in fact, represents the {\it quantum uncertainty} in 
momentum (in units of $M_s$) 
of the recoiling D-particle. 

The spacetime (\ref{metricrecoil}) 
may be considered as a {\it mean field} result of appropriate
resummation of quantum corrections for the collective coordinates
of the recoiling $D$-particle. To lowest order in a weak 
string ($g_s < 1$) $\sigma$-model perturbative framework, 
such quantum fluctuations may be obtained by resumming 
pinched annuli world-sheets, which can be shown 
to exponentiate~\cite{szabo}, thereby providing a 
Gaussian probability distribution 
over which one averages.
The metric (\ref{metricrecoil}) is the result of such an average,
and the eventual identification of the Liouville mode with the target 
time~\cite{emn}.

For low energies compared to 
string scale $M_s$, the parameter $b'=b'(E)$ is given by~\cite{szabo}:
\ba
b'(E)=4g_s^2\Bigl(1 - \frac{285}{18}g_s^2\frac{E}{M_D}\Bigr) 
\label{bdef} 
\ea
where $E$ denotes the {\it kinetic} energy of the recoiling 
$D$-particle, 
$g_s$ is the string coupling, assumed weak, 
and $M_D=M_s/g_s$ is the D-particle mass, which is formally 
derived in the logarithmic conformal field theory approach 
from energy-momentum conservation~\cite{szabo}. 

It should be remarked at this point that we shall be working 
throughout this work with $g_s $ small but {\it finite}. 
The limit 
$g_s \rightarrow 0$ will not be considered, given that when $g_s =0$ 
the mass of the D-particle $M_D \rightarrow \infty$, and thus
the recoil is absent, but on the other hand the curvature
of the surrounding spacetime as a result of the immense mass
of the $D$-particle should be taken into account. In that limit
the scale $b'(E) \rightarrow 0$, and thus one can no longer 
consider distances sufficiently far away from the center of the infinite 
gravitational attraction so that the Schwarzschild curvature
effects of the $D$ particle could be ignored.

As one observes from (\ref{bdef}), the value of $b'$
decreases with increasing energy, and formally vanishes 
when the energy is close to $M_s$. 
We should note, however, that the above expression (\ref{bdef})
pertains strictly to slowly moving strings, i.e. $E \ll M_s$.
In general, for arbitrary energies (including intermediate ones, which we shall
be interested in below), the precise 
expression for $b'(E)$ is not known at present.
For our purposes, 
we shall assume that $b'(E)$ decreases with increasing
$E$ for all energies. This will be 
justified later on.

A crucial ingredient of the approach is 
the identification of target time with the Liouville mode $\phi$~\cite{emn}. 
In order for this procedure to be consistent, the resulting 
effective field theory must satisfy the appropriate $\sigma$-model 
conformal invariance conditions, which in a target-space framework 
correspond to appropriate equations of motion derived from a 
string-effective action. In this note we shall 
demonstrate that this is indeed the case, to lowest order 
in a Regge slope $\alpha '$ perturbative expansion. 
We shall also study some physically important properties 
of the spacetime (\ref{metricrecoil}).
We shall argue that this spacetime is unstable, with 
the inevitable result of emission of high-energy radiation. 
Then we shall 
speculate on possible 
astrophysical applications of this phenomenon. 
Specifically, we shall argue that, as a result of the 
high-energy-photon emission from the unstable bubble, the neighborhood of the 
recoiling D-particle defect may behave as a source of ultra-high-energy 
particles, which can reach the observation point, provided that the defect
lies at a distance from Earth which is within the respective mean-free paths
of the energetic particles. This effect, then, may 
have
a potential connection with the recently observed~\cite{gzkobs} apparent 
`violations' of Greisen-Zatsepin-Kuzmin (GZK) cutoff~\cite{gzkcutoff}.

\section{Dynamics of the Recoil spacetime} 

In four-dimensional spacetime, obtained by appropriate compactification
of the higher-dimensional spacetime, where string theory lives, 
the string massless multiplet 
consists of 
a graviton field $g_{\mu\nu}$, a dilaton $\Phi$
and an antisymmetric tensor field $B_{\mu\nu}$, which in four dimensions
gives rise, through its field strength $H_{\mu\nu\rho}=\partial_{[\mu}B_{\nu\rho]}$, to a pseudoscalar axion field $b$ (not to be confused
with the uncertainty parameter $b'$). The latter is defined as follows:
\ba
H_{\mu\nu\rho}=\frac{1}{\sqrt{-g}}\epsilon_{\mu\nu\rho\sigma}\partial_\sigma b
\label{axion}
\ea 
What we shall argue below is that the spacetime (\ref{metricrecoil}) 
is compatible with the equations of motion obtained from a string  
effective action for the above fields. Equivalently, these 
equations are the $\sigma$-model conformal invariance conditions
to leading order in the Regge slope $\alpha'$.
We stress once again the fact that the spacetime metric 
(\ref{metricrecoil}) has been  deived upon the non-trivial assumption
that the target time is identified with the Liouville field~\cite{emn},
whose presence is necessitated by the recoil~\cite{recoil,kogan,szabo}.
The fact, as we shall see, that this identification is consistent
with the $\sigma$-model conformal invariance conditions 
to ${\cal O}(\alpha')$,
is therefore a highly-non-trivial consistency check of the approach.

The components of the Ricci tensor for the metric (\ref{metricrecoil}) are:
\ba
R_{00}=-\frac{2{b'}^2}{t^2}~, \qquad R_{ij}=\frac{\delta_{ij}}{r^2}-\frac{x_ix_j}{r^4} ~, \qquad i,j=1,2,3. 
\ea
The curvature scalar, on the other hand, reads:
\ba
R=-\frac{4}{r^2} 
\ea
which is independent of $b'(E)$ and singular at 
the origin $r=0$ (initial position of the D-particle). 
Thus, we observe that the spacetime after the recoil acquires
a singularity. However, 
our analysis is only valid for distances $r$ larger than the 
Schwarzschild radius of the massive $D$-particle, and hence 
the locus of points $r=0$ cannot be studied at present
within the perturbative $\sigma$-model appropach. It is therefore 
unclear whether the full stringy spacetime has a true singularity 
at $r=0$. 

The Einstein tensor $G_{\mu\nu} \equiv R_{\mu\nu}-\frac{1}{2}g_{\mu\nu}R$
has components:
\ba
G_{00}=0~, \qquad G_{ij}=-\frac{\delta_{ij}}{r^2}-\frac{x_ix_j}{r^4}
\ea
The conformal invariance conditions for the graviton mode 
of the pertinent $\sigma$-model 
result in the following Einstein's equations as usual:
\ba
G_{\mu\nu}=-{\cal T}_{\mu\nu} 
\ea
where ${\cal T}_{\mu\nu}$ contains contributions from string matter, 
which in our case includes 
dilaton and 
antisymmetric tensor (axion) fields, and probably cosmological constant
terms (which will turn out to be zero in our case, as we shall see later on).
As we shall also show, there are tachyonic modes necessarily present,
which, however, are not the ordinary flat-spacetime Bosonic 
string tachyons. In fact,
despite the fact that so far we have dealt explicitly with bosonic actions, 
our approach is straightforwardly extendable 
to the bosonic part of  superstring
effective actions. In that case, ordinary tachyons are absent from the
string spectrum. However, our type of tachyonic modes, will still be present
in that case, because as we shall argue later, such modes
simply indicate an instability of the spacetime (\ref{metricrecoil}). 

We find it convenient to use a redefined stress energy tensor 
${\cal T}'_{\mu\nu} \equiv {\cal T}_{\mu\nu}-\frac{1}{2}g_{\mu\nu}{\cal T}_\alpha^\alpha$,
in terms of which Eisntein's equations become:
\ba\label{einst} 
R_{\mu\nu}=-{\cal T}'_{\mu\nu} 
\ea
The stress tensor ${\cal T}'_{\mu\nu}$ for the case of tachyon and 
axion fields reads: 
\ba
{\cal T}'_{\mu\nu}= \partial_\mu T \partial_\nu T 
+ \partial_\mu b \partial_\nu b - g_{\mu\nu}V(T) 
\label{stress}
\ea
where $V(T)$ is a potential for the tachyonic mode $T$. 
The fact that the axion field $b$ 
does not have a potential is dictated by the abelian gauge symmetry 
of string effective actions, according to which they depend only on 
the antisymmetric-field strength $H_{\mu\nu\rho}$ and {\it not} on the 
field $B_{\mu\nu}$. 
Below we shall show that indeed the field $T$
acquires a tachyonic mass, which however, in contrast to the flat-space
time Bosonic string theory tachyons, depends on the parameter $b'(E)$.

In addition to (\ref{einst}), one has the conformal invariance 
conditions for the tachyon and axion fields, to ${\cal O}(\alpha ')$:
\ba 
\partial^2 T = -V'(T)~, \qquad \partial^2 b = 0 
\label{axiontachyon}
\ea
where the prime denotes differentiation with respect to $T(x_i,t)$.  

\begin{figure}
\epsfxsize=3.5in
\bigskip
\centerline{\epsffile{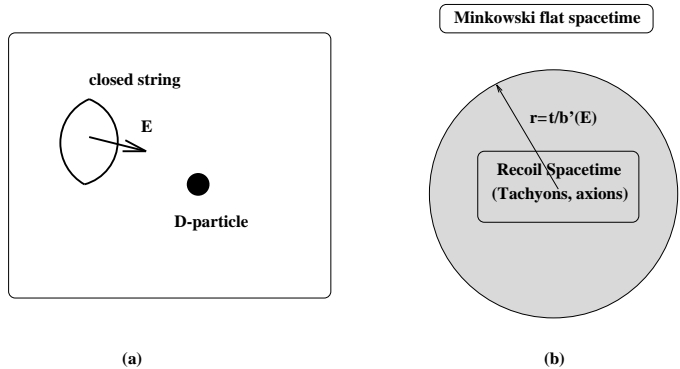}}
\vspace{0.2in}
\caption{(a) Scattering of a closed string mode 
(of energy E and non-trivial momentum) off a D-particle 
embedded in a four-dimensional spacetime (obtained by compactification 
of a higher-dimensional stringy spacetime). (b) after the recoil: 
formation of effective bubble, 
in the interior of which
the tachyonic modes 
and axion fields are confined. The  exterior geometry is the 
flat Minkowski.
\label{fig:bubble}}
\end{figure}

A solution to (\ref{einst}), (\ref{axiontachyon}) is given by:
\ba
T(x_i,t)= {\rm ln}r~, \qquad b(x_i,t)=b'(E){\rm ln}t 
\label{solution}
\ea
provided that the tachyon potential $V(T)$ is:
\ba
V(T) = -{\rm exp}\Bigl(-2T(r)\Bigr)
\ea
Naively, if the solution is extended to all space, we observe that 
the matter diverges logarithmically. To remedie this fact
we restrict the above 
solution to the range $r \le t/b'(E)$, and thus we enclose it in a 
{\it bubble} of time-dependent radius $t/b'(E)$. 
Outside the bubble we demand the spacetime to be the flat Minkowski
spacetime, and thus the above upper limit in $r$, $t/b'(E)$  
is the locus of points
at which the temporal component of the metric (\ref{metricrecoil}) 
becomes unity, and this allows an appropriate matching of the interior
and exterior geometries. 
As we shall see later on, 
a non-trivial consistency check of this matching 
will be provided {\it dynamically} 
by an explicit study of the scattering of 
test particles off the bubble. 
In this way the phenomenologically 
unwanted tachyon and probably axion fields (obtained
from the antisymmetric tensor field of the string) are 
confined inside the bubble of radius $t/b'(E)$ (cf. figure \ref{fig:bubble}).

In arriving at the above solution we have restricted ourselves to 
${\cal O}(\alpha')$ because we have ignored terms of higher order
in curvature $R$. To justify such an approximation it suffices to 
note that the ratio of the leading terms to the next to leading ones 
is:
\ba\label{ho} 
(R/M_s)^2/R = {\cal O}\Bigl(b^2\Bigr)={\cal O}\Bigl(g_s^2(1-
g_s^2\frac{285}{18}\frac{E}{M_D})\Bigr)
\ea
provided one does not 
approach the singularity at $r=0$, which is in fact consistent 
with the regime of validity of the logarithmic conformal field 
theory~\cite{kogan,szabo}, i.e. 
distances far away from the defect, 
and times long enough after its collision with the string. 
This implies that our analysis is restricted near the 
boundary of the bubble, which will be 
sufficent for our purposes in this work. 
Since the string coupling is assumed weak, it 
is evident from 
(\ref{ho}) that the approximation of neglecting the higher-curvature
terms is satisfactory near the boundary of the bubble. Then, 
from the decreasing behaviour of $b'(E)$ with increasing energy, which, as 
mentioned previously, is assumed here even for intermediate energies, 
it follows that this approximation becomes 
even better for higher energy scales appropriate for the early stages 
of the universe.

A second important remark concerns the fact that in
the analysis leading to the metric (\ref{metricrecoil}) 
we have treated the spacetime surrounding the D-particle defect
as initially (i.e. before the collision with the string) flat.
However, even the initially at rest $D$-particle, being a very massive one of mass $M_D=M_s/g_s$ 
would naturally curve the spacetime around it, 
producing a Schwarzschild radius
$r_S=\ell_P^2/g_s\ell_s$, where $\ell_P$ denotes the 
four-dimensional Planck length 
and $\ell_s=1/M_s$ the string length. 
From our discussion above, the radius of the bubble of figure \ref{fig:bubble}
is $r_b=\ell_s/g_s $. For consistency of our approach, the approximation of 
treating the spacetime as initially flat implies that 
we work in distances considerably larger than the Schwarzschild radius,
so as the general relativistic effects due to the mass of the $D$-particle 
could be safely ignored. This implies that the radius of the bubble 
must be considerably larger than $r_S$, i.e. 
\ba
   1 \gg r_s/r_b = \Bigl(\ell_P/\ell_s\Bigr)^2  
\ea
{}From the modern view point of string/$D$-brane theory, the string length 
may not be necessarily of comparable order as the Planck legth,
but actually could be larger. Thus, the above condition seems consistent. 

Notice also that the matching with the flat Minkowski spacetime in the 
exterior geometry is possible because the matter energy density 
${\cal T}_{00}$  and energy flow ${\cal T}_{0i}$ 
in the interior of the bubble of figure \ref{fig:bubble} are {\it both zero}:
\ba
{\cal T}_{00}={\cal T}_{0i}= 0 
\ea
and thus there is no radiation coming out of or flowing into the bubble. 

We next compute the mass squared of the field $T$.
To this end, we shall consider the fluctuations $\delta T$ 
of the tachyonic mode 
$T$ around the classical solution $T_{\rm cl}$ 
(\ref{solution}) in the interior 
of the bubble, but close to its boundary, where the spacetime 
approaches the Minkowski flat geometry. From (\ref{axiontachyon}) 
we then have: 
\ba
\partial^2 \delta T - 4e^{-2T_{\rm cl}}\delta T \Bigl|_{r\rightarrow t/b'(E)}=0 
\ea
from which we obtain a mass squared term of the form:
\ba\label{mass} 
  m^2 =-4\frac{{b'}^2(E)}{t^2}~. 
\ea 
The negative value indicates, of course, the fact that the field $T$ is 
tachyonic, but the interesting issue here is that the induced mass
depends on the parameter $b'(E)$, and hence on the initial energy 
data of the incident string. 
 
A remark we would like to make at this point concerns the time-dependence
of the mass (\ref{mass}). As one observes from (\ref{mass}), the 
tachyon field will eventually disappear from the spectrum 
(as it becomes massless) asymptotically in time $t$. This fact comes from the 
specific form of the metric (\ref{metricrecoil}). 
However, as argued in \cite{recoil} quantum effects will eventually
stop the expansion of the bubble and may even force it to contract.
In this sense, the value of the mass of the tachyonic mode (\ref{mass}) 
will remain finite and negative, and will never relax to zero. 

An equivalent way of seeing this is to observe that 
the time $t$ in the temporal component of the metric 
(\ref{metricrecoil}) may be absorbed by a redefinition of the time 
$t \rightarrow t'= {\rm ln}t$. In that case, the metric reads:
\ba
   ds^2={b'}^2(E)r^2 d{t'}^2 - \sum_{i=1}^{3} dx_i^2
\ea
Under this redefinition, the bubble solution remain, but this time the 
bubble appears to be independent of time, with its radius being 
$r=1/b'(E)$, and the mass squared of the tachyon being $m^2=-4{b'}^2$.
From that we observe that the tachyon mass remains 
$b'(E)$-dependent and finite. 
This argument supports the fact that even in the initial 
coordinate system (\ref{metricrecoil}) the time $t$ cannot be such so as 
to eliminate the $b'(E)$ dependence of the tachyon mass. 
This system of coordinates corresponds to 
a frame in which the bubble appears static, hence it corresponds 
in some sense to a \emph{comoving frame}. Therefore the mass
$m^2=-4{b'}^2(E)$ is the rest mass of the tachyonic mode. 

Some clarifying remarks are in order at this point concerning the 
nature of the tachyonic mode $T$. As we have just seen its mass is dependent
on the uncertainty scale $b'(E)$ of the D-particle, and hence is proportional
to the string coupling $g_s$ (cf. (\ref{bdef})). 
For this reason this tachyonic mode should be distinguished 
from the standard tachyon fields in flat-spacetime free Bosonic string theory. 
In fact, in our approach this tachyonic mode expresses simply
the {\it instability} of the {\it bubble} configuration, 
and will be present even in superstring effective field theories. 

As a result, from such consideration one may obtain an average
{\it lifetime} $\tau$ for the bubble:
\ba\label{lifetime} 
   \tau = \frac{1}{2b'(E)}
\ea

In our approach we 
prefer to work with the initial form of the metric 
(\ref{metricrecoil}), implied directly by the logarithmic
conformal field theory approach to recoil~\cite{kogan,recoil},
which defines a natural frame for the definition of the observable time.
In this approach, the presence of the recoil degrees of fredom 
after a time, say, $t=0$ imply a breaking of the general coordinate 
invariance by the background, and also an irreversibility of 
time~\cite{recoil}. This comes from the fact that in our problem, time is 
a world-sheet renormalization group parameter (Liouville mode)~\cite{emn}, 
which 
is assumed irreversible, flowing towards a non-trivial infrared fixed
point 
of the world-sheet renormalization group 
of the Liouville 
$\sigma$-model~\cite{zam,recoil}. In this frame the mass of the tachyon
appears time dependent, because, as we shall discuss below, the frame 
is not an inertial one, given that the spacetime of the bubble is a Rindler
\emph{accelerated spacetime}.

\begin{figure}
\epsfxsize=2.5in
\bigskip
\centerline{\epsffile{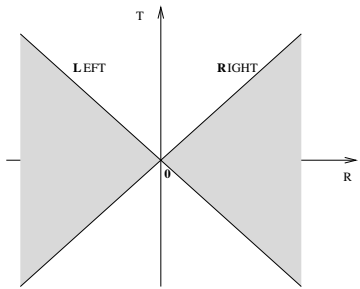}}
\vspace{0.2in}
\caption{Rindler wedge spacetime arising from the 
recoil of a $D$-particle, embedded in a 
four-dimensional spacetime, due to its scattering
with a closed string. The set of wedges (LEFT and RIGHT)
describe the spacetime for $t>0$. 
\label{fig:rindler}}
\end{figure}

\section{Thermal Effects and Radiation from the Bubble Spacetime}

To show this we consider 
the metric (\ref{metricrecoil}), pass onto spherical polar coordinates $(r,\theta,\phi)$, and fix 
the angular part for convenience, as this does not modify 
the concusions. We then 
perform the following coordinate transformations (from now on we work 
in units $\ell_s=1$):
\ba\label{rindlertrnsf} 
R=r{\rm cosh}(b'(E){\rm ln}t)~, \qquad T=r{\rm sinh}(b'(E){\rm ln}t),
\ea
The transformation maps our space time to the right Rindler wedge (R) depicted
in figure \ref{fig:rindler}. The left wedge (L) is described 
by similar transformations up to a minus sign. 
In the $(R,T)$ coordinates the line element becomes:
\ba
ds^2=dT^2-dR^2 - (R^2-T^2)d\Omega ^2 
\ea
where $d\Omega$ is the conventional solid angle.

From the above spacetime, we observe that for distances $R \gg T$ 
one recovers the flat Minkowski spacetime, whilst for distances
$R \sim T$ one obtains the {\it bubble} spacetime. 
Notice that $R^2-T^2=r^2$, and the interior of the bubble is defined 
by $r \le 1/b'(E)$ in comoving coordinates. 

An observer comoving with the expanding bubble, placed at position $r$,
is accelerated with respect to the $(R,T)$ frame, with proper 
accelration $1/r$. According, then, to the standard analysis 
of accelerated observers~\cite{davies}, such an observer sees
the Minkowski vacuum (in the $(R,T$) coordinates)
as having a non-trivial {\it temperature} $T_{\rm bubble}$  
\ba 
    T_{\rm bubble}=\frac{1}{2\pi r}, \qquad  0 < r < 1/b'(E)
\ea
The temperature for the inertial Rindler observer $T_{0}$
is:
\ba\label{trindler}
T_0 =\sqrt{g_{00}}T_{\rm bubble}=\frac{b'(E)}{2\pi}
\ea
The presence of temperature is expected to imply a non-trivial proper 
entropy density $s$. For a massless scalar field, in our case the axion $b$, 
the entropy $S=\int d^3x \sqrt{-g} s$ 
in four spacetime dimensions is given by: 
\ba
  S=\int d^3x \sqrt{-g} \frac{4}{3}\frac{\pi^2}{30}T_{0}^3 
\ea
From the bubble spacetime (\ref{metricrecoil}) one then obtains:
\ba\label{entropy} 
S=\int d^3x \sqrt{-g} s=4\pi b'(E) \int _0^{1/b'(E)}dr~r^3  
\frac{4}{3}\frac{\pi^2}{30}\frac{{b'}^3(E)}{8\pi^3}  =\frac{1}{180} 
\ea
Thus the bubble carries non-trivial entropy, which turns out to 
be independent 
of $b'(E)$. The reader should not be alarmed by the apparent volume 
independence of the entropy, which at first sight 
would seem to contradict the fact 
that the entropy is an extensive quantity. In fact, there is no 
contradiction in our case, since 
there is only one scale in the problem, 
$b'(E)\ell_s$, and the volume of the bubble is itself expressed
in terms of this scale.

The presence of entropy production after the recoil implies 
{\it loss of information} which can be understood as follows: 
one starts from a pure state of a string striking a $D$ particle. 
There is no entropy in the initial configuration. 
After the strike, the $D$-particle recoils, and because it is a 
heavy object it distorts the spacetime around it, 
producing the bubble phenomenon via 
its recoil excitation degrees of freedom.
Due to the finite lifetime of the bubble, 
the entropy (\ref{entropy}) will be released to the external space,
implying information encoded in the recoil degrees of freedom, which 
are unmeasurable by an asymptotic observer. 
This picture is consistent with the loss of conformal invariance
of the underlying $\sigma$-model, and the irreversible 
world-sheet renormalization-group 
flow of the recoiling system, as discussed in \cite{recoil,emn}, 
upon the identification of the Liouville mode with the target time. 

It should be remarked at this point 
that the entropy (\ref{entropy}) 
pertains strictly to scalar fields that live {\it 
inside the bubble}. There is no crossing of the surface of the bubble 
by the interior fields in our construction, at least classically
(as we shall discuss below there is a quantum-mechanical 
escape probability). 
On the other hand, it should be 
noticed that for a field 
which lives in the {\it exterior} of the bubble, 
there appears to be 
loss of information in the sense 
that the exterior particle degrees of freedom 
may enter the bubble and be captured, as we shall discussed 
in the next section. 
Thus, 
an asymptotic
observer, far away from the bubble, will necessarily {\it trace out}
such (unobserved) 
degrees of freedom in a density matrix formalism, and in this 
sense the resulting entropy,
pertaining to such degrees of freedom,
will be proportional 
to the {\it area} of the bubble and {\it not its volume}~\cite{srednicki}.
We shall come back to this important point, in connection with 
emitted radiation from the bubble later on. 

\begin{figure}
\epsfxsize=2.5in
\bigskip
\centerline{\epsffile{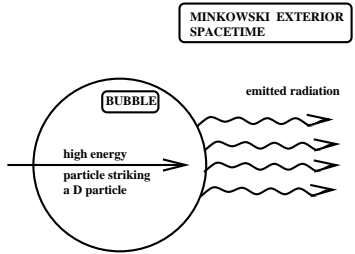}}
\vspace{0.2in}
\caption{Emitted radiation from the unstable 
bubble. The radiation is not isotropic, 
but most of it will be emitted in the forward direction,
parallel to that of the incident high-energy particle.
\label{fig:thermal}}
\end{figure}

The presence of temperature $T_0$ (\ref{trindler}) implies the 
emission of radiation from the bubble (see figure \ref{fig:thermal}), 
which can be 
read off from the Stefan-Boltzman law $\sigma T_0^4$, $\sigma = \pi^2/60$ 
(in units $\hbar=c=k_B=1$). Given that the area is $4\pi {b'}^{-2}(E)$,
and that the lifetime of the bubble is estimated from (\ref{lifetime}),
one observes that during the life time $\tau$, the following 
amount of energy is released in the form of radiation:
\ba\label{energyrad} 
    E_{{\rm rad}} \sim 4\pi {b'}^{-2}(E) \tau \sigma T_0^4 =
\frac{1}{480\pi}b'(E)M_s=\frac{g_s}{480\pi}b'(E)M_D  
\ea 
It is interesting to observe that the same amount of energy 
represents the {\it thermal} energy of the axion field in the interior 
of the bubble:
\ba
  E_{{\rm th/axion}}=\int d^3x \sqrt{-g}\frac{\pi^2}{30}T_0^4 
= \frac{g_s}{480\pi}b'(E)M_D
\ea
From (\ref{bdef}), and taking into account that in the effective 
field-theory 
limit we are working here, $E < M_s$, and that $g_s \ll 1$ 
in our weak string framework, 
one observes that the energy $E_{{\rm rad}}=E_{{\rm th/axion}} < M_D$.

From energy conservation then, which, notably, 
is shown to be valid rigorously 
in the context of our logarithmic conformal field theory 
(stringy) recoil framework~\cite{szabo},
one obtains:
\ba
E_{\rm in} + M_D = M_D + E + E_{\rm th/axion} 
\ea
where $E_{\rm in}$ denotes the total energy of the {\it incident} 
particle/string. 
From this, one thus sees that there is a {\it threshold} 
for bubble formation:
\ba\label{threshold} 
E^{\rm threshold} = E_{\rm rad}= E_{\rm th/axion}
\ea
From these considerations, 
one observes that the radiation energy {\it will not cause any 
mass loss} of the $D$-particle, since all the thermal axion energy
accounts for that. Hence, despite the instability 
of the bubble, the stability of the D-particle defect is not affected.
This will be important for our physical applications, to be discussed 
later on.

Notice that the energy release (\ref{energyrad}) 
cannot make up for the 
maximum of the thermal energy expected from Wien's law 
$\lambda T_{\rm max} = {\rm const}$, where $\lambda$ the wavelength
of radiation. Thus the resulting photon spectrum is {\it not thermal},
and hence one can only get an estimate for the energies of the 
emitted radiation. The alert reader might then object to our
previous use of the black body (or, in general, equilibrium) laws.
In fact their use is indicative, and they can only give
qualitative results (e.g. in our case we only obtain a  
tail of the thermal distribution).

To recapitulate, the physics behind the above properties
can be summarized as follows: one needs a highly-energetic 
incident particle of energy $E_{\rm in}> E^{\rm threshold}$, which strikes
a $D$-particle, and forms a bubble; the bubble radiates an amount 
of energy 
$E^{\rm rad}$ (\ref{energyrad}) distributed appropriately among 
the various photons. The emitted radiation will not be isotropic 
as a result of (spatial) momentum conservation
(see figure \ref{fig:thermal}). This will be useful 
in physical applications. 
The fact that a particle, entering 
and being captured by the bubble, will
cause the 
emission of radiation from the bubble
is nicely related to the existence of non-zero entropy 
measured by an asymptotic observer. 
This phenomenon is thus not dissimilar to the Hawking process of an
evaporating black hole, although in our case 
the bubble is {\it not} a black hole, neither the D-particle evaporates,
as we discussed above. This picture is in agreement 
with the 
non-equilibrium Liouville string framework~\cite{recoil}, on which  
the approach is based.

\section{Motion of particles in the bubble spacetime}

We shall now analyze the motion of a particle 
in the bubble spacetime (\ref{metricrecoil}). 
For convenience we shall work at the equator of the three sphere 
(fixed angle $\theta =\pi/2$). The lagrangian of the particle 
in the background spacetime (\ref{metricrecoil}) in comoving coordinates 
$(t',r, \phi)$ is:
\ba
{\cal L}=\frac{1}{2} \Bigl(\frac{ds}{d\lambda}\Bigr)^2 =
\frac{{b'}^2(E)r^2}{2}\Bigl(\frac{dt'}{d\lambda}\Bigr)^2 
- \frac{1}{2}\Bigl(\frac{dr}{d\lambda}\Bigr)^2 - r^2\frac{1}{2}
\Bigl(\frac{d\phi}{d\lambda}\Bigr)^2 
\ea
Expressing the Lagrangian in  terms of the conserved angular momentum 
$L=r^2 (d\phi/d\lambda)^2$
and energy $E_{\rm in}={b'}^2(E)r^2 \tfrac{dt'}{d\lambda}$ we obtain:
\ba
\frac{E_{\rm in}^2}{2{b'}^2(E)r^2}-\frac{1}{2}\Bigl(\frac{dr}{d\lambda}\Bigr)^2 
- \frac{L^2}{2r^2} =\frac{\mu^2}{2}
\ea
where 
$\mu=0$ for massless particles (e.g. photons), and $1$ for massive particles
(in which case the quantities $L$ and $E_{\rm in}$ 
are the corresponding quantities per 
unit mass).

Writing the equation of motion as:
\ba\label{effpot}
  \frac{1}{2}\Bigl(\frac{dr}{d\lambda}\Bigr)^2+\frac{\mu^2}{2}  =
\frac{E_{\rm in}^2/{b'}^2(E) - L^2}{2r^2}
\ea
we can see that the impact parameter 
$L/E_{\rm in}$ must be smaller than $1/b'(E)$, for the equation to make sense.
This means that if $L/E_{\rm in} > 1/b'(E)$, then the particle 
will {\it necessarily} travel {\it outside} the bubble spacetime, 
which is thus a dynamical consistency check of our matching 
assumptions that the spacetime in the region $r \ge 1/b'(E)$ 
is the flat Minkowskian spacetime. Such outside particles 
will not be affected by the presence of the bubble, and their 
trajectory will be undisturbed, that predicted by special 
relativistic dynamics.

Below we shall study now the case of impact parameters 
$L/E_{\rm in} < 1/b'(E)$,
for massless and massive particles. In such a case, from (\ref{effpot}) 
the massless particle equation of motion reads: 
\ba \label{plusminus} 
   r=r_0
{\rm exp}\Bigl(\pm\phi \sqrt{\frac{E_{\rm in}^2}{{b'}^2(E)L^2}-1}\Bigr)
\ea
where in the case of an incoming photon (from outside the bubble) 
$r_0=\frac{1}{b'(E)}$, and we have taken only the 
$(-)$ sign, because this is the only consistent choice. 
This shows that the massless particle will be
{\it captured} inside the bubble. 

From (\ref{effpot}) one observes that, 
in the case of a massive particle, there exist values of energy and 
angular momentum such that the particle can escape the bubble spacetime. 
This happens if 
the 
radial velocity on the boundary (as the particle attempts to escape)
is non-zero, which implies (we have re-instated the units of $M_s$ 
for clarity):
\ba\label{escapecond} 
E_{\rm in}^2 - L^2{b'}^2(E)M_s^2 > M_s^2
\label{condition} 
\ea
This demonstrates that only highly energetic particles with energies
much higher than $M_s$ can escape the bubble spacetime.

\begin{figure}
\epsfxsize=2.5in
\bigskip
\centerline{\epsffile{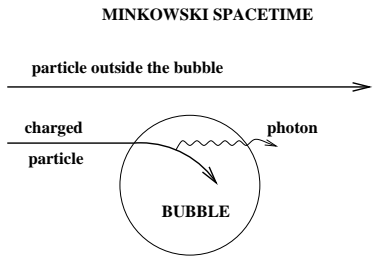}}
\vspace{0.2in}
\caption{Emission of photons due to non-uniform motion 
of a charged particle inside the bubble. The radiation 
can escape the bubble.
\label{fig:transition}}
\end{figure}

Once such a particle is electrically charged, its non-uniform 
(spiral) motion inside the bubble will  cause the emission of 
{\it radiation}. The latter will continue to carry angular momentum 
and energy 
of roughly the same order as that of the particle. Because 
the emission is now taken place at $r_0 < 1/{b'}(E)$ 
the positive sign in the exponent of equation (\ref{plusminus}) 
is also allowed, implying an escape possibility for the emitted 
photons (cf. figure \ref{fig:transition}). 
In addition, as we shall show in the next section, the 
bubble has a non-trivial (thermal) refractive index, and thus 
behaves as a medium. If there is a beam of charged particles 
entering the bubble within its short life time, then 
these particles will experience the (suppressed) phenomenon 
of {\it transition radiation}~\cite{trt}, i.e. the emission 
of photons accompanying 
an electrically-charged particle when  
crossing the interface separating two media with different
refractive indices (in our case, the interior of the bubble and the 
exterior Minkowski spacetime). 
A fraction of this radiation  
will also escape the bubble. 
Such phenomena, if true, will imply excess of photons
accompanying the charged particle. We shall present a more 
detailed discussion of these issues, and their 
potential experimental consequences, 
in a forthcoming publication.

It should be noticed that, if the condition (\ref{condition}) is satisfied, 
then the particle is {\it deflected} by an angle $\Delta\phi$ 
which can be computed in a standard way to be:
\ba\label{deflection}  
\Delta\phi = \pi - \frac{2}{\sqrt{\frac{1}{\rho^2{b'}^2(E)}-1}}
{\rm arccosh}\Bigl(E_{\rm in}\sqrt{1-\rho^2{b'}^2(E)}\Bigr)~, \qquad \rho=L/E_{\rm in}={\rm impact~parameter}
\ea
{}From this we observe that, for fixed impact parameter $\rho$, 
highly energetic particles will not be deflected much, as 
should be expected.

As a final comment we mention that the scattering 
cross section $\sigma (E)$ 
for energies and/or angular momenta 
that violate the 
condition (\ref{condition}), which are of physical interest, 
is given by:
\ba\label{crossection} 
\sigma (E) = \pi {b'}^{-2}(E) 
\ea
From the results of ref. \cite{szabo}, which are valid strictly 
only for low energies,
we observe that $b'(E)$ decreases with increasing energy $E$ (\ref{bdef}). 
One would expect intuitively that strings with higher energy 
would cause larger distortion of the spacetime surrounding the 
recoiling $D$-particle defect. This point of view is 
supported by  the above results if one 
extends the behaviour of 
$b'(E)$ encoded in (\ref{bdef}) to intermediate 
energies as well, and in fact to all energies (up to Planckian), 
because in that case the distortions of spacetime 
caused by strings with higher energy will correspond to 
formation of bubbles with bigger radii $1/b'(E)$. 

From (\ref{crossection}) we observe that 
higher energy strings would correspond to larger cross sections.
The important point to notice is that the cross section 
is non zero even for zero energy. This stems from (\ref{bdef}),
and is associated with the fact that $b'(E)$ is 
essentially a  {\it quantum uncertainty} 
in the momentum of the $D$-particle~\cite{szabo}, which is 
not zero even for vanishing incident energy $E$.  
Hence, the spatial uncertainty of the position of the 
$D$-particle, which is associated with the cross section, is not zero, 
but is bounded from below by the Heisenberg uncertainty principle,
which explains naturally the non-zero minimum value of $\sigma (E=0)$. 
It also explains the increasing behaviour of the 
cross section with increasing energy, given that the higher 
the energy is, the larger the uncertainty is expected to be. 

\section{Quantum Electrodynamics Effects inside the Bubble and 
Refractive Index for Photons}

So far our considerations have been classical. 
It is in this sense that we demonstrated capture of 
photons (and, in general, massless particles) 
in the interior of the bubble. 
The presence of 
finite temperature (\ref{trindler}) will create a 
non-trivial (thermal) vacuum, with broken Lorentz symmetry 
inside the bubble. In that case, it is known~\cite{pascual}
that the effective velocity of light, defined by the 
quantum propagator of photons, is modified in accordance 
with the fact that the finite temperature effects 
provide the notion of a medium. 

Specifically, the dispersion relation for a 
particle of mass $m$ in the non-trivial 
vacuum at temperature $T$ is: $E^2(p)=p^2 + f(p,T,m)$,
where $E$ is the energy and $p$ the momentum, and 
the function $f$ encodes the quantum effects of the 
vacuum polarization. The group velocity of the particle 
is then given by $v=\partial E(p)/\partial p$, and in general 
depends on $p$. 

In the case of photons in a 
non-trivial quantum electrodynamical vacuum, the function 
$f$ can be computed, to one loop, from the vacuum-polarization graph 
of the photon~\cite{pascual}. The latter
is of order $f \sim T^2e^2$, where $e$ is 
the electron charge~\footnote{Note that one can extend these results to 
incorporate all known particles in the standard model~\cite{pascual}.}. 

In our case, the induced temperature (\ref{trindler}) is much larger than the 
incident momentum $p$ of the photons. From the results of \cite{pascual}
in this case, the effective velocity of the photons 
inside the bubble is (in units of speed of light {\it in vacuo} $c=1$):
\ba\label{index} 
v \sim \frac{p}{eT_0} \sim \frac{p}{e {b'}(E)}   
\ea
where $p$ is related to the energy $E$ of the photon 
via the (modified) dispersion relation. The photon in this case
becomes {\it effectively} massive, with mass $\mu \sim e {b'}(E) \ne 0$. 
It is interesting to note that, 
due to the extreme temperature effects, $p \ll T_0$, the resulting photon 
is considerably slowed down inside the bubble 
to non-relativistic velocities.

From our earlier discussion on massive particle trajectories
inside the bubble, and the 
fact that the quantum effects result in an effective photon mass,
one is tempted  to consider the
possibility of a photon escaping the bubble. 
However, 
from the condition (\ref{escapecond}) 
and the induced photon mass $\mu \sim e {b'}(E)$, 
it becomes evident 
that 
there is {\it no such possibility}.

Nevertheless, there is a non-zero 
probability of {\it quantum tunneling} 
through the potential barrier.  
In this sense, part of the classically captured (incident) photons can 
escape.
Combined with the thermal slowing down (cf. (\ref{index})) of all photons 
inside the bubble, then, 
this will result in the appearance of 
{\it delays} in the respective arrival times of photon beams
from distant astrophysical sources in areas where there are 
D-particles. Because these delays will be associated with 
only part of the photon beam, the final effect will appear 
as a fluctuation in the arrival time.

The associated delay for a single photon, which   
passes through a region in space 
where there is one D-particle, and is assumed to escape 
through tunelling, 
is estimated to be: 
\ba\label{delay} 
   \Delta t \sim \frac{1}{b'(E)v(E)} = \frac{e}{p} 
\ea
where $\frac{1}{b'(E)}$ is the radius of the bubble, $e$ is the 
electron charge,
and this formula is applicable for $p \ll M_s$. For velocities
$p \sim M_s$ there are modifications, which however 
are not of interest
to us here. 

The existence of delay effects that depend on the energies of the photons
bares some resemblance to quantum spacetime 
effects, associated with induced refractive index for photons, diuscussed in 
\cite{sarkar}. As in those works, so in the present model there 
appears to be a non-trivial refractive 
index $n(E)=1/v(E)$ inside the bubble. However, 
there is an important difference, in that
here this is due to conventional quantum electrodynamical thermal effects. 
Because of this reason, the induced refractive index (\ref{index}) 
is reduced with increasing photon energies, in contrast to the 
effects of refs. \cite{sarkar}, where the quantum spacetime induced 
refractive index appears to increase with energy. 
However, the reader 
should bear in mind that in our model the existence of temperature is due to 
quantum stringy effects~\cite{recoil,szabo} associated with the 
recoil of the $D$-particle. In this sense there is, 
in our approach, a notion of quantum gravity, in similar spirit 
to the work of \cite{sarkar}.  

Unfortunately, for a very dilute gas of D-particles the maximal 
delay 
(\ref{delay}), corresponding to the less energetic 
observable photons (e.g. infrared background of 
energies 0.025 eV) is very small, or order $10^{-11}$ s.
Nevertheless, we should notice that if one considers
photon sources from distant astrophysical objects that 
are at distances corresponding to cosmological redshifts $z > 1$,
then in those areas of the relatively early universe, the density 
of D-particles might have been higher. 
One then can get an upper bound of such densities 
by considering the effect (\ref{delay})
as lying inside the present experimental 
errors. We hope to come to a more systematic 
analysis of such effects in a forthcoming publication.

\section{Discussion and possible physical applications on the GZK cutoff}

In this work we have studied the formation of bublles
as a result of scattering of closed strings off D-particles
embedded in a four dimensional spacetime. 
Such configurations may be thought of as a trivial case of 
intersecting branes, provided one adopts the modern view point
that our four-dimensional world is a $D3$-brane. 
In this sense, the string scale $M_s$, which enters our 
calculations, is not necessarily the same with the 
four-dimensional Planck scale, and as we have seen this played
an important r\^ole in our analysis, as it allowed us to 
work at distances sufficiently far from the Schwarzschild radius
of the D-particle, for finite $g_s$ string couplings.

An interesting feature of our approach 
is the entropy production, which we associated
with information carried away by the recoil degrees of freedom. 
This latter feature may have cosmological implications for 
mechanisms of entropy production in the early universe, where 
we expect the density of D-particles to be significant, and hence 
the probability of scattering with closed strings important.
The fact that highly energetic charged particles
can escape the bubble, with the simultaneous release of radiation,
which can also escape and thus is in principle 
observable, is interesting, and might 
imply important phenomenological constraints on the 
order of the density of the D-particles in the Universe today.
We plan to present a systematic study of such issues in a 
forthcoming publication. 

Another comment we wish to make concerns the 
impossibilty of the extension of the above 
analysis, with the specific choice of fields,
to higher-dimensional 
spacetimes, of spacetime dimension $d>4$.
This comes about because, under the 
simplest form for the tachyonic mode $T = {\rm ln}r$,
consistent with   
Einstein's equations 
to order ${\cal O}(\alpha')$,  
one obtains 
the following form for the tachyonic-mode potential
$V(T) \sim 1/r^2$.  
On the other hand, the 
equation of motion for the mode itself, consistent 
with the above form for $T$,  
demands that $dV(T)/dT =-(d-2)V(T)$. 
Clearly, this is consistent  
only {\it in d=4} spacetime dimensions. 
Hence, despite the fact that in our approach we have 
restored Lorentz invariance, 
we still obtain a special r\^ole of $d=4$, 
as in the 
Lorentz violating scenario of \cite{recoil},
where the sacrifice of Lorentz invariance in the 
sense of an explicitly 
space-dependent vacuum energy,
lead also to a selection of $d=4$.
If this feature survives the inclusion of the complete string 
matter multiplets, something which is not clear to us at present, 
then, it might imply that a {\it recoiling} 
Liouville $D$-particle cannot be embedded in (intersect with)
a target spacetime
(viewed itself as a D-brane), consistently 
with the $\sigma$-model conformal invariance, 
unless 
its dimensionality is $d=4$. At present we consider the issue 
only as a mathematical curiosity of the specific
effective field theory at hand, and 
we do not attribute it further physical 
significance. However, surprises cannot be excluded. 

A final comment we would like to make concerns possible physical 
applications of the above phenomena. 
The emitted radiation from the bubble spacetime may find
important astrophysical applications as can be seen by the following
simplified scenario: there is a rare distribution of D-particles
in the inflated universe (although their density was much higher in the early universe). Due to this rare distribution, the concept of isotropy 
is not applicable. It is therefore possible that an isolated D-particle
lies between Earth and a distant galaxy.
As the galaxy emits particles, some highly energetic 
and weakly interacting ones, such as neutrinos, strike the 
D-particle and induce the formation 
of a bubble. The latter then emits radiation in the way 
explained above. The emitted photons, in the direction of the incident 
particles, will be highly energetic, of typical energy $b'(E)$. 
In general, such photons will interact with the background photons 
of either the microwave background radiation, or the infrared background,
to yield, say, $e^+e^-$ pairs. Because of this, if the D-particle 
lies far away from Earth, outside the average mean free path 
of such photons, the latter will not arrive on Earth. 
However, one may imagine a situation in which the isolated D-particle
lies within the above mean free path distance, which in the case 
of photons interacting with the infrared background 
is estimated to be of order of a few Mpc~\cite{protheroe}. 
Then, the weakly interacting incident particles, that trigger the phenomenon
of bubble formation, e.g. 
neutrinos, will  
arrive at the location of the D-particle(s) undisturbed. 
In that case, the 
emitted high-energy radiation will reach the observation point, and 
in this sense the recoiling D-particle bubble {\it constitutes a novel 
and relatively nearby source of highly energetic photons}. Such scenaria 
may have applicability to the recently observed 
highly energetic 30 TeV photons that seem to violate the so-called
GZK cut-off~\cite{protheroe,gzkobs}. 

In a similar way, one can also extend the above discussion to incorporate
charged particles. Consider, for instance, a beam of protons 
emitted by a galaxy lying at cosmological distances, 
whose energies
do not exceed the GZK cut-off, and hence they arrive 
undisturbed
until the point where a D-particle defect lies.
One of the protons will then strike the D-particle
and create a bubble spacetime. As discussed previously, the proton
will be captured inside the bubble, since its 
energy does not satisfy the escape condition (\ref{escapecond}), 
as being less than $M_s$~\footnote{Here we assume that 
$M_s$ exceeds the GZK cut-off $10^{19} {\rm eV} = 10^{10}$ GeV. 
If this is not the case, and one has a lower $M_s$, then such 
massive particles may escape. At any rate, 
our discussion does not 
depend upon this fact.}. 
The bubble will then radiate very high energy photons,
which can interact with the remaining protons 
in the beam, that fly outside the bubble, to create, say, 
protons and pions {\it etc}. The protons (or, in general, 
the particles) that emerge from 
such interactions   
will then be very energetic, and it is conceivable that their energies
can be of the order of the observed~\cite{gzkobs} $3 \times 10^{20}$ eV.
In this way, the region around the recoiling D-particle defect
acts as a source of ultra-high-energy cosmic rays, and if one 
assumes, as before, that the defect lies within the mean free path 
of a proton (from Earth), this can easily explain the 
observed apparent ``violations'' of the GZK cut-off~\cite{gzkobs}. 

If the above scenario survives, it may then 
imply that there is no actual violation of Lorentz symmetry 
that is responsible for the  phenomenon, as claimed by a number 
of authors~\cite{gzk}, since in our bubble spacetime 
there is no such violation (except the trivial one due to temperature
effects inside the bubble). To put it in simple terms, 
in our scenario, the source of the ultra-high-energy 
cosmic rays is in the neighborhood of the D-particle defect, 
which may lie at a much closer distance from Earth 
than one naively thought. A more detailed discussion 
of these speculative scenaria 
will constitute the topic of a forthcoming 
publication. 
Of course, it goes without saying that one cannot exclude 
the possibility that 
a peculiar  
combination of phenomena, involving the model discussed here
in conjunction with both conventional
and unvonventional (spacetime foam) physics~\cite{sarkar,gzk}, 
might 
actually
lie behind such extreme astrophysical phenomena. 

Admittedly, the above ideas are very speculative, 
and indeed may have nothing to do at the end with  
real Physics. However, the mathematical and logical consistency
of such unconventional effects seems, at least to the authors, 
convincing enough so as not to discard them immediately. 
It is actually very intriguing that stringy defects,
which at first sight are not expected to play any r\^ole 
in low-energy physics, may actually be responsible 
for some extreme astrophysical phenomena, whose observation 
became only recently possible, as a result of the 
enormous technological advances in 
both terrestrial and extraterrestrial instrumentation.  
It is always intellectually challenging, but also expected intuitively 
in some sense, to think of the vast Universe as 
being the `next-generation'
Laboratory, where ideas on the quantum structure of spacetime
may be finally subjected to experimental tests in the not-so-distant future.

\section*{Acknowledgements} 

The work of E.G. is supported 
by a King's College London Research Studentship (KRS).
N.E.M. wishes to thank H. Hofer (ETH, Zurich and CERN) 
for his interest and partial support.

\end{document}